\documentclass{iaus}
\usepackage{graphicx}
\usepackage{natbib}

\title[R4.2~~Large 3D kinematical surveys - high z] 
{3D spectroscopic surveys: Exploring galaxy evolution mechanisms}

\author[Beno\^it Epinat]   
{Beno\^it Epinat$^{1,2}$
}

\affiliation{$^1$ Universit\'e de Toulouse; UPS-OMP; IRAP; Toulouse, France\\
$^2$ CNRS; IRAP; 14, avenue \'Edouard Belin, F-31400 Toulouse, France\\
email: {\tt benoit.epinat@ast.obs-mip.fr}
}

\pubyear{2010}
\volume{277}  
\pagerange{1--6}
\jname{Tracing the ancestry of galaxies on the land of our ancestors}
\editors{C. Carignan,K. C. Freeman, \& F. Combes, eds.}

\def\jnl@style{\it}
\def\aaref@jnl#1{{\jnl@style#1}}

\def\aaref@jnl#1{{\jnl@style#1}}

\def\aj{\aaref@jnl{AJ}}                   
\def\araa{\aaref@jnl{ARA\&A}}             
\def\apj{\aaref@jnl{ApJ}}                 
\def\apjl{\aaref@jnl{ApJ}}                
\def\apjs{\aaref@jnl{ApJS}}               
\def\ao{\aaref@jnl{Appl.~Opt.}}           
\def\apss{\aaref@jnl{Ap\&SS}}             
\def\aap{\aaref@jnl{A\&A}}                
\def\aapr{\aaref@jnl{A\&A~Rev.}}          
\def\aaps{\aaref@jnl{A\&AS}}              
\def\azh{\aaref@jnl{AZh}}                 
\def\baas{\aaref@jnl{BAAS}}               
\def\jrasc{\aaref@jnl{JRASC}}             
\def\memras{\aaref@jnl{MmRAS}}            
\def\mnras{\aaref@jnl{MNRAS}}             
\def\pra{\aaref@jnl{Phys.~Rev.~A}}        
\def\prb{\aaref@jnl{Phys.~Rev.~B}}        
\def\prc{\aaref@jnl{Phys.~Rev.~C}}        
\def\prd{\aaref@jnl{Phys.~Rev.~D}}        
\def\pre{\aaref@jnl{Phys.~Rev.~E}}        
\def\prl{\aaref@jnl{Phys.~Rev.~Lett.}}    
\def\pasp{\aaref@jnl{PASP}}               
\def\pasj{\aaref@jnl{PASJ}}               
\def\qjras{\aaref@jnl{QJRAS}}             
\def\skytel{\aaref@jnl{S\&T}}             
\def\solphys{\aaref@jnl{Sol.~Phys.}}      
\def\sovast{\aaref@jnl{Soviet~Ast.}}      
\def\ssr{\aaref@jnl{Space~Sci.~Rev.}}     
\def\zap{\aaref@jnl{ZAp}}                 
\def\nat{\aaref@jnl{Nature}}              
\def\iaucirc{\aaref@jnl{IAU~Circ.}}       
\def\aplett{\aaref@jnl{Astrophys.~Lett.}} 
\def\apspr{\aaref@jnl{Astrophys.~Space~Phys.~Res.}}
\def\bain{\aaref@jnl{Bull.~Astron.~Inst.~Netherlands}} 
\def\fcp{\aaref@jnl{Fund.~Cosmic~Phys.}}  
\def\gca{\aaref@jnl{Geochim.~Cosmochim.~Acta}}   
\def\grl{\aaref@jnl{Geophys.~Res.~Lett.}} 
\def\jcp{\aaref@jnl{J.~Chem.~Phys.}}      
\def\jgr{\aaref@jnl{J.~Geophys.~Res.}}    
\def\jqsrt{\aaref@jnl{J.~Quant.~Spec.~Radiat.~Transf.}}
\def\memsai{\aaref@jnl{Mem.~Soc.~Astron.~Italiana}}
\def\nphysa{\aaref@jnl{Nucl.~Phys.~A}}   
\def\physrep{\aaref@jnl{Phys.~Rep.}}   
\def\physscr{\aaref@jnl{Phys.~Scr}}   
\def\planss{\aaref@jnl{Planet.~Space~Sci.}}   
\def\procspie{\aaref@jnl{Proc.~SPIE}}   

\begin{document}

\maketitle

\begin{abstract}

I review the major surveys of high redshift galaxies observed using integral field spectroscopy techniques in the visible and in the infrared. The comparison of various samples has to be done with care since they have different properties linked to their parent samples, their selection criteria and the methods used to study them.
I present the various kinematic types of galaxies that are identified within these samples (rotators, mergers, etc.) and summarize the discussions on the mass assembly processes at various redshifts deduced from these classifications: at intermediate redshift ($z\sim 0.6$) merger may be the main mass assembly process whereas the role of cold gas accretion along cosmic web filaments may increase with redshift. The baryonic Tully-Fisher relation is also discussed. This relation seems to be already in place 3 Gyr after the Big-Bang and is then evolving until the present day. This evolution is interpreted as an increase of the stellar mass content of dark matter haloes of a given mass.
The discovery of positive abundance gradients in MASSIV and LSD/AMAZE samples is highlighted. At $z\sim 3$ this discovery might be linked to cold gas accretion along cosmic filaments toward the centre whereas at lower redshift ($z\sim 1.3$), this may be mainly due to accretion of gas from outer reservoirs toward the centre via tidal tails due to interactions.

\end{abstract}

\firstsection 
\section{Introduction}

In the $\Lambda$CDM framework, haloes form from the initial density fluctuations of the Universe. They grow in a hierarchical way from merging of various small mass haloes. Inside these haloes, the baryonic matter is collapsing and forms the first stars in the first galaxies. Thus, in this framework, merging is a natural mechanism to explain the mass assembly of galaxies. However, recent cosmological numerical simulations have highlighted that another process might be efficient at high redshift: fresh cold gas might be accreted along cosmic filaments onto galaxies (e.g. \citealp{Dekel:2009}).
These processes might be responsible for the building of the modern day Hubble sequence.
In order to study the contribution of these processes along the cosmic time, several high redshift galaxy surveys using 3D spectroscopy techniques are being studied in various redshift ranges. From these data, galaxy mass assembly processes are mainly studied through galaxy kinematics. Some studies are also focused on gas chemical abundances.
The main aim of this review is to present and compare the results of these studies.

In section \ref{samples}, I present the main samples and their selection criteria. In section \ref{variety}, I discuss the galaxy variety observed in each sample. In section \ref{tullyfisher}, I focus on the evolution of the Tully-Fisher relation with redshift and finally, in section \ref{abundance}, I report results obtained from studies of abundance gradients in high redshift galaxies.


\section{Samples}
\label{samples}

The main samples of 3D spectroscopy data at high and intermediate redshifts are focused in the redshift range $0.4<z<4$. This is the period during which the morphological transition between irregular high redshift galaxies and regular morphologies as seen in the modern Universe is occurring. In addition, the Universe is experiencing a peak of cosmic star formation activity for $1<z<2$. This star formation decreases after $z\sim 1$.


These samples are observed using integral field spectroscopy in the visible or in the infrared depending on (i) the redshift of the sources and (ii) the emission lines that are targeted to study the galaxy physical properties.
The IMAGES sample \citep{Yang:2008,Neichel:2008,Puech:2008,Rodrigues:2008}
is focused in the redshift range $0.4<z<0.75$. The \textsc{[Oii]} doublet is observed in the visible with \textsc{Flames/Giraffe} spectrograph on the VLT.
At higher redshift, three samples are observed using the near-infrared spectrograph \textsc{Sinfoni} on the VLT to target primarily the H$\alpha$ line:
(i) the MASSIV sample (\citealp{Epinat:2009,Queyrel:2009}; Contini et al. in prep; Epinat et al. in prep; Vergani et al. in prep; Queyrel et al. in prep) in the redshift range $0.9<z<1.8$, (ii) the SINS sample \citep{Genzel:2008,Shapiro:2008,Shapiro:2009,Cresci:2009,Forster-Schreiber:2009} in the redshift range $1.3<z<2.7$ and (iii) the LSD/AMAZE sample \citep{Maiolino:2008,Mannucci:2009,Gnerucci:2010,Cresci:2010} in the redshift range $2.5<z<4.0$.
Two other smaller samples have been observed using the near-infrared spectrograph OSIRIS on the Keck telescope: one at $1.5<z<1.7$ \citep{Wright:2009} and the other at $2.0<z<2.5$ \citep{Law:2009}.


\begin{table}
\begin{center}
\begin{tabular}{|c||c|c|c|c|}
\hline
 Sample & $z$ & Size & Parent sample & Selection \\
\hline
\hline
\textbf{IMAGES} & 0.4 -- 0.75 & 63 & CDFS + others & $EW([O\textsc{ii}])\geq15\AA$, $M_J(AB)\leq-20.3$ \\
\hline
\textbf{MASSIV} & 0.9 -- 1.8 & 85 & VVDS &  $EW([O\textsc{ii}])\geq30\AA$ or UV slope\\
\hline
\textbf{SINS} & 1.3 -- 2.7 & 80 & Multiple & LBG (color) \\
\hline
\textbf{LSD/AMAZE} & 2.5 -- 4.0 & 32 & From \citet{Steidel:2003} & LBG (color) \\
\hline
\textbf{OSIRIS} & 1.5 -- 1.7 & 6 & From \citet{Steidel:2004} & LBG (color) \\
\hline
\textbf{OSIRIS} & 2.0 -- 2.5 & 13 & Multiple & Multiple + line flux\\
\hline
\end{tabular}
\caption{Samples properties}
\label{samples_table}
\end{center}
\end{table}

\subsection{Selection}

The comparison of these various samples needs caution due to their different selection criteria.
Such samples can be drawn from several parent samples. These parent samples are built from large spectroscopic surveys.
Several criteria are used in order to select star forming galaxies. Some criteria are related to line measurements on the spectroscopic data that might be used as proxies for the star formation rate of galaxies: equivalent widths or line fluxes. They also usually combine magnitude limits linked to the completeness of the parent samples.
Other criteria are related to the colours of galaxies in order to select lyman break galaxies for instance or to ensure that there exists a population of young stars thus tracing star formation (cf. Table \ref{samples_table}).

In order to understand the effect of these selection criteria, the representativeness of the samples with respect to the underlying population of galaxies in each redshift bin has to be studied.
For example, the MASSIV sample is mainly missing very massive galaxies of the whole galaxy population as probed by the VVDS sample \citep{LeFevre:2005} in the redshift bin $0.9<z<1.8$, from which MASSIV has been built. MASSIV is representative of VVDS star-forming galaxies although the median [O\textsc{ii}] flux and equivalent width are slightly higher than for the whole parent sample
(Contini et al. in prep).
On the other hand, the SINS sample has not been built from a unique sample. However, in order to assess the representativeness of this sample, \citet{Forster-Schreiber:2009} have compared it to the Chandra Deep Field South (CDFS) in the redshift range $1.3<z<2.6$ because it is one of the best-studied deep survey fields. It results that the selection criteria used for SINS lead to the selection of the massive star forming galaxies of the galaxy population.

\subsection{The data}

Even if the redshift range is large, within a standard cosmology, the physical scale only varies slightly (between 6 kpc/arcsec and 8.5 kpc/arcesc). Since the data are mainly seeing limited, the spectral resolution is not strongly depending on the instrumentation used. The typical resolution is around 6 kpc. The gain in AO data is a factor from 2 to 4 in spatial resolution ($\sim 2$ kpc).
The spectral resolution of the various instruments is around 3000 in the infrared and above 10000 in the visible (\textsc{Flames/Giraffe}).

Since galaxies are large enough to be observed as extended sources using 3D spectroscopy, one limitation for AO observations is that the sensitivity is low due to the small pixel size. In addition, extended sources are subject to surface brightness dimming inducing a decrease of the surface brightness with redshift proportional to $(1+z)^4$.
Lensed surveys offer a good opportunity to target the faintest sources at high redshift with a good spatial resolution since lensing induces both spatial and flux magnifications. However, such samples are limited to few targets \citep{Jones:2010}.

From 3D spectroscopy data, various maps are built: line flux map, velocity field, velocity dispersion map. For high redshift galaxies, these maps are affected by beam smearing: for rotators, when beam smearing increases, the velocity gradient decreases and a central velocity dispersion peak grows \citep{Epinat:2010}. However the latter effect is less visible when galaxies have intrinsically high velocity dispersions. Similar beam smearing corrections are used in each sample described here in order to infer the rotation velocity and the local gaseous velocity dispersion of galaxies.

%
%
%



\section{Galaxy variety}
\label{variety}

The kinematics enables a diagnostic on the nature of the galaxies. Are galaxies rotating regularly? Are their kinematics perturbed? Are they in a merging state? Are they dominated by rotation or dispersion? etc.
The dynamical nature of galaxies is used to infer the mass assembly processes in action.

\subsection{Classification}

Various kinematics classifications have been proposed in the different surveys.
The classification proposed in the frame of IMAGES takes advantage of the low spatial resolution to propose three classes based on the position of the velocity dispersion peak and on the velocity field regularity \citep{Yang:2008}:
rotating disks when a velocity dispersion peak is observed at the centre, perturbed rotators when the peak is offset from the centre and complex kinematics when the peak if offset and the velocity field irregular.
From this classification, \citet{Yang:2008} found a large fraction of galaxies with anomalous kinematics and concluded that the rapid evolution of galaxy kinematics from $z\sim0.6$ and $z=0$ may be mostly induced by merging.

In the frame of MASSIV (Epinat et al. in prep), on the one hand, galaxies are classified according to their kinematics to distinguish galaxies with high velocity shear ($\Delta V >50$ km/s) and low velocity shear ($\Delta V <50$ km/s). Around one third of the first epoch MASSIV sample has a low shear. This fraction can not be explained only by disks seen face on. However, some stable disks with regular kinematics and low local velocity dispersion (similar to local disks) have been found in that sample.
On the other hand, close galaxy environment leads to a second classification: when counterparts are detected both in H$\alpha$ and in continuum images galaxies are classified as in interaction. It results that around one third of MASSIV galaxies are in interaction, mainly minor ones. These results imply that several processes might still be acting at $z\sim1.3$ in contrast with $z\sim 0.6$.

The classification of the SINS survey is based on a kinemetry analysis of both the velocity field and the velocity dispersion map \citep{Shapiro:2008}. A diagnostic diagram based on the asymmetry of these kinematic maps is built in order to separate rotators and mergers. A second classification is based on the dynamical support to separate rotation-dominated and dispersion-dominated galaxies.
From the whole sample, \citet{Forster-Schreiber:2009} found one third of dispersion-dominated disks interpreted as resulting from a cold gas accretion mechanism as well as a significant fraction of mergers (one third). In the same redshift range, \citet{Law:2009} found non rotating objects from OSIRIS data that they interpreted as resulting from cold gas accretion.

From these studies, a coherent scenario can be built but a consensus is not reached on the interpretation of the kinematics. However, it is clear that at high redshift, there exists disks that have a much larger gaseous velocity dispersion than in the local Universe.

%
%

\subsection{Clumpy galaxies: which processes are responsible?}

At high redshift, many galaxies look clumpy. At $z\sim 2$, these clumpy galaxies are associated to turbulent rotating disks (SINS, \citealp{Genzel:2010}). Both these morphological and kinematic features are expected in a cold gas accretion scenario which might be efficient at these redshifts (e.g. \citealp{Bournaud:2009}). Thus, these clumpy galaxies are thought to be formed via a cold gas accretion mechanism.

At lower redshift ($z\sim 0.6$), \citet{Puech:2010a} observed that galaxies from the IMAGES sample with clumpy morphologies have a lower velocity dispersion than at higher redshift. Since half of these galaxies are compatible with major mergers and because cold gas accretion might not be efficient at this redshift \citep{Keres:2009}, he concludes that interactions are the main driver for clump formation at this epoch.

\section{Baryonic Tully-Fisher relation evolution}
\label{tullyfisher}

\begin{figure}[t]
\begin{center}
 \includegraphics[width=4.5cm]{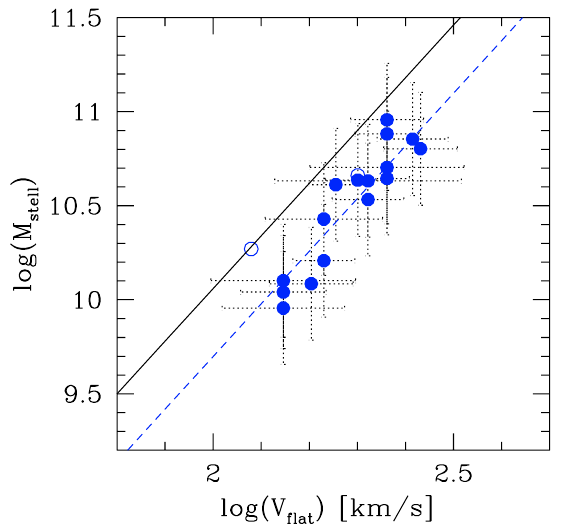}
 \includegraphics[width=4.3cm]{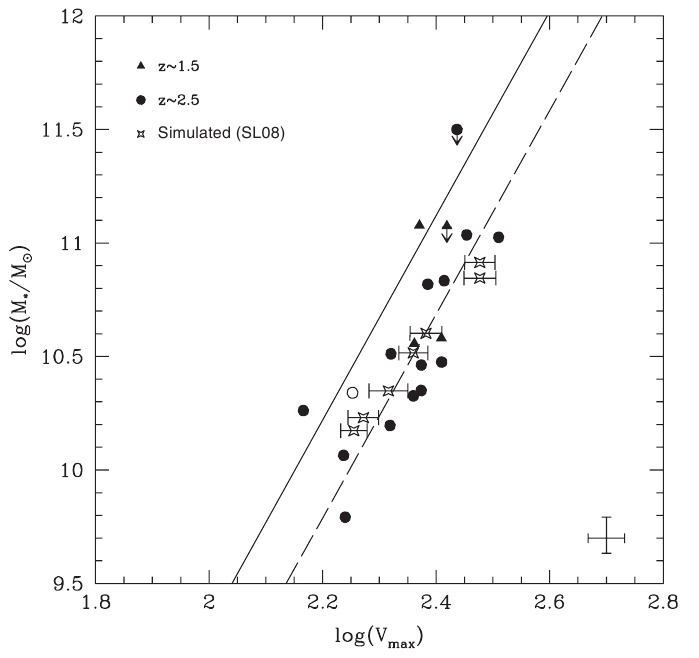}
 \includegraphics[width=4.1cm]{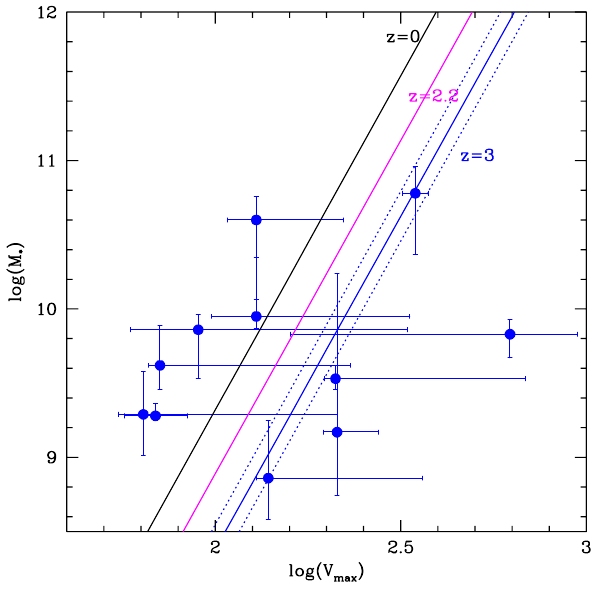}
   \caption{Stellar mass Tully-Fisher relation: \textit{Left:} IMAGES sample, $z\sim 0.6$ \citep{Puech:2008}. \textit{Middle:} SINS sample, $z\sim 2.5$ \citep{Cresci:2009}. \textit{Right:} LSD/AMAZE sample, $z\sim 3.0$ \citep{Gnerucci:2010}. Black solid lines represents the $z=0$ relation. The same reference is used in SINS and LSD/AMAZE. The dotted and/or the blue lines represent the fit to the data for each sample forcing the slope to the local one.}
   \label{fig1}
\end{center}
\end{figure}

Originally, the Tully-Fisher relation (TFR) was used in the local Universe as a distance estimator \citep{Tully:1977}. This relation links rotational velocity of disk galaxies with their magnitude. In other words, it links their total mass content to their stellar content. At higher redshift, due to galaxy evolution, changes in this relation are expected if one assumes that dynamical mass is dominated by halo mass and that haloes and stellar content are not evolving at the same rate. In particular, galaxies host more and more stars while the Universe evolves.
At high redshift, people study primarily the stellar TFR, using stellar masses derived from SED fitting. The study of the baryonic TFR would need a constraint on the gas content which is still difficult at the present time at high redshift due to the sensitivity of current radio telescopes. However, the first observations of the molecular gas are being obtained in massive high-redshift galaxies  \citep{Bothwell:2010,Tacconi:2008,Tacconi:2010}.

Within the various high redshift samples from which is studied the TFR, the authors restrict their analysis on galaxies identified as regular rotators. This reduces the scatter found in their relation. They find a monotonic evolution of the TFR zero point from $z=0$ up to $z\sim 3$ (cf. Figure \ref{fig1}): (i) from IMAGES, \citet{Puech:2010a} found an evolution compatible with a doubling of the stellar mass in a given halo mass during the last 6 Gyr; (ii) from MASSIV, Vergani et al. (in prep) found a TFR compatible with the one at $z\sim 0.6$ but with a slightly larger scatter; (iii) from SINS, \citet{Cresci:2009} found a small scatter and an evolution compatible with an increase of the stellar mass by a factor 2.5 between $z\sim 2.2$ and $z=0$; (iv) from LSD/AMAZE, \citet{Gnerucci:2010} observed that the TFR was already in place at $z\sim 3$ and found a stronger evolution but the statistics is poor and the scatter is large.

The comparison of these samples has to be modulated by the various sample selections that could in particular explain the large scatter observed in some cases. In addition, the samples of rotators are relatively small in each study (less than 20 galaxies). One other issue concerns the local Tully-Fisher relations used as reference in these studies that have quite different slopes. In the local Universe, there appears to exist a break, i.e. various slopes depending on the mass range of galaxies. At high redshift, the various samples have quite different stellar masses properties making the comparison non trivial.


\section{Abundance gradients}
\label{abundance}

Recently, the first positive abundance gradients have been observed in high redshift galaxies \citep{Cresci:2010}: metallicity maps of three galaxies at $z\sim3$ from the LSD/AMAZE sample classified as regular rotationally supported disks were derived using a combination of three independent metallicity diagnostics.
Due to the nature of these galaxies, the authors concluded that these unexpected positive gradients may be generated by cold flows of fresh gas along cosmic filaments toward the centre.

At lower redshift, in the frame of the MASSIV sample, Queyrel et al. (in prep) also found seven galaxies with secure positive metallicity gradients deduced from the ratio [N\textsc{ii}]/H$\alpha$. The majority of these galaxies (5/7) are associated with  interacting/merging systems.
Thus, these behaviours might be interpreted as fresh gas accretion in the centre due to interaction tidal tails.

\section{Conclusions and perspectives}
\label{perspectives}

From the analysis of 3D spectroscopic surveys of high redshift galaxies, there seems that cold gas accretion is a more efficient mass assembly mechanism in the early Universe than at later epochs where merging might be the main mechanism. Evidence for this statement are: (i) the existence of a large population of high velocity dispersion disks at $z>1$ that is no more observed at $z\sim 0.6$; (ii) the different nature of clumpy galaxies at $z\sim 2$ and $z\sim 0.6$; (iii) the different origin for metallicity gradients at $z\sim 2$ and $z\sim 1.3$.
The smooth evolution of the TFR  with redshift is coherent with a scenario in which galaxies in a given halo mass are increasing their stellar content with time. However, in order to know if gas content is already in place in these haloes it will be necessary to study neutral and molecular gas content with new radio telescopes such as ALMA or SKA (and precursors). These instruments will be well suited as well to study high redshift galaxy kinematics. 
New instruments like MUSE/VLT or KMOS/VLT will also be well adapted to build large, deep and diverse samples. At longer term, the Extremely Large Telescopes will enable to reach a very high spatial resolution in 3D spectroscopy studies of high redshift galaxies and thus probe in details the physical mechanisms at play in the early assembly of galaxies. 

%
%
%

\begin{acknowledgements}
Part of this work (the MASSIV project) is supported by the french ANR grant ANR-07-JCJC-0009, the 
CNRS-INSU and its Programme National Cosmologie-Galaxies.
\end{acknowledgements}

\bibliographystyle{aa}
\bibliography{biblio}

%
%
%

\end{document}